\documentclass{jetpl}
\twocolumn

\lat

\title{Superconducting energy gap distribution in c-axis oriented MgB$_{2}$
thin film from point contact study}

\rtitle{Superconducting energy gap distribution in MgB$_{2}$}

\sodtitle{Superconducting energy gap distribution in c-axis oriented MgB$_{2}$
thin film from point contact study}

\author{Yu.\,G.\,Naidyuk\/\thanks{e-mail: naidyuk@ilt.kharkov.ua }, I.\,K.\,Yanson,
L.\,V.\,Tyutrina, N.\,L.\,Bobrov, P.\,N.\,Chubov,
W.\,N.\,Kang$^{*}$, Hyeong-Jin Kim$^{*}$, Eun-Mi Choi$^{*}$, and
Sung-Ik Lee$^{*}$ }

\rauthor{Yu.\,G.\,Naidyuk, I.\,K.\,Yanson, L.\,V.\,Tyutrina et
al.}

\sodauthor{Yu.\,G.\,Naidyuk$^{+}$, I.\,K.\,Yanson$^{+}$, L.\,V.\,Tyutrina$^{+}$
et. al.}

\address{B.Verkin Institute for Low Temperature Physics and
Engineering, National Academy  of Sciences of Ukraine, 47 Lenin
Ave., 61103,  Kharkiv, Ukraine\\~\\
$^*$National Creative Research Initiative Center for
Superconductivity, Department of Physics, Pohang University of
Science and Technology, Pohang 790-784, South Korea}

\dates{15 January 2002}{*}

\abstract{We have analyzed about hundred  voltage-dependent
differential resistance $dV/dI(V)$ curves of metallic point
contacts between $c$-axis oriented MgB$_{2}$ thin film and Ag,
which exhibit clear Andreev reflection features connected with the
superconducting gap. About one half of the curves show the
presence of a second larger gap. The histogram of the double gap
distribution reveals distinct maxima at 2.4 and 7 meV, while
curves with a single-gap features result in more broad maximum at
3.5 meV. The double-gap distribution is in qualitative agreement
with the distribution of gap values over the Fermi surface
calculated by H. J. Choi et al. (cond-mat/0111183). The data
unequivocally show the presence of two gaps $\Delta _{S}=2.45\pm
0.15$ meV and $\Delta _{L}=7.0\pm 0.45$ meV in MgB$_{2}$ with gap
ratio $\Delta _{L}/\Delta _{S}=2.85\pm 0.15$. Our observations
prove further a widely discussed multigap scenario for MgB$_{2}$,
when two distinct gaps are seen in the clean limit, where  a
single averaged gap is present in the dirty one.}


\begin{document}

\maketitle

\textit{Introduction}.
 Direct spectroscopic investigations of the superconducting
order parameter in recently discovered \cite{Nagam} superconductor
MgB$_2$ with T$_c\simeq$40\,K by tunneling
\cite{Karapet,Rubio,Sharoni,Chen,Giubileo1,Giubileo2,Badr,Zhang,Zasad}
and point-contact spectroscopy
\cite{Zasad,Deuts,Plec01,Laube,Gonelli,Szabo,Bugos,Bobrov} show
unambiguously an energy gap $\Delta $ in the quasiparticle density
of states (DOS). However, the experimental results are
controversial as to the gap width $\Delta $, whose variation from
1.5 to 8 meV (see, e. g., review \cite{Buzea}) is unexpectedly
large, pointing to the possibility of multiphase  or
nonhomogeneous samples, degraded surface, or anisotropic  energy
gap. Another way to solve this puzzle is to consider two
superconducting gaps in MgB$_{2}$, as proposed by Liu {\it et al.}
\cite{Liu}, accounting complex electronic structure of MgB$_{2}$
with both quasi-2D and 3D Fermi surface \cite{Kortus}. Indeed,
several papers \cite{Giubileo1,Giubileo2,Badr,Szabo,Bugos,Bobrov}
have reported double gap structure in the differential conductance
(resistance) with the smaller gap being far below weak-coupling
BCS value $\Delta $=1.76k$_{\rm B}$T$_c\simeq $6\,meV and the
larger gap slightly above the standard BCS one, in accordance with
theory \cite{Liu}.

Therefore, one of the intriguing key issues of superconducting
state of MgB$_2$ is whether the double gap state is intrinsic or
the spread of the gap values is a result of anisotropy,
nonhomogeneity, surface effect, etc. In other words, before
macroscopic high quality single crystals will be available for
thorough investigations, the sample imperfection may raise doubts
about the final conclusion.  However, in our mind, good
reproducibility of the double-gap values given by different
authors \cite{Giubileo1,Giubileo2,Badr,Szabo,Bugos,Bobrov} by
different, in their physical background, methods such as tunneling
and point-contact spectroscopy carried out on different samples
such as pellets, films, grains, all this with a great degree of
probability supports intrinsic nature of the double gap in
MgB$_2$.

In this paper we will give further confirmation of double gap scenario
in MgB$_2$ based on analysis of about hundred point-contact
spectra of c-axis oriented thin films.

\textit{Experimental and calculation details}.
 We have measured the high-quality c-axis oriented 0.4\,$\mu$m
thick MgB$_2$ film  \cite{Lee01} grown by a PLD technique on Al$_2$O$_3$
substrate. The resistivity of the film exhibits a
sharp transition at 39 K with a width of $\sim$ 0.2 K from 90\% to
10\% of the normal state resistivity \cite{Lee01}. The residual
resistivity $\rho_0$ at 40\,K is $\sim$ 6 $\mu\Omega$\,cm
\footnote{There is a scattering by factor of 4 in $\rho_0$ for the
different films.} and RRR=2.3.

Different point contacts (PCs) were established in situ directly
in liquid $^{4}$He by touching  as-prepared surface (sometimes
etched by 1\% HCl solution in ethanol) of the MgB$_2$ film by a
sharpened edge of an Ag counterelectrode, which were cleaned by
chemical polishing in HNO$_3$. This geometry corresponds to the
current flowing preferably along the c axis. A number of contacts
were measured by touching of the film edge after breaking
Al$_2$O$_3$ substrate. By this means, the current flows preferably
along the ab plane. The differential resistance d$V/$d$I$ vs $V$
was recorded using a standard lock-in technique. The normal
resistance $R_N$ (at $V\gg\Delta$) of investigated contacts ranged
mainly between 10 and 1000 $\Omega$ at 4.2\,K.

The important characteristic of  PC is their size or diameter $d$,
which can be determined from the simple formula derived by Wexler
\cite{Wexler} for contact resistance:
\begin{equation}
\label{Rwex} R_{\rm PC}(T) \simeq  \frac {16 \rho l}{3\pi d^2} +
\frac{\rho (T)}{d},
\end{equation}
where two terms represent ballistic Sharvin \footnote{In the case
of interface scattering Sharvin resistance should be multiple by
factor (1+Z$^2$)\cite{BTK82}.} and diffusive Maxwell resistance,
correspondingly. Here $\rho l = p_{\rm F}/n$e$^2$, where $p_{\rm
F}$ is the Fermi momentum and $n$ is the density of charge
carriers. The latter for MgB$_2$ is estimated at $n\simeq
6.7\times 10^{22}$ \cite{Canfield}, which results in $\rho l\simeq
2\times 10^{-12}\Omega\cdot $cm$^2$ using $v_{\rm F}\simeq 5\times
10^{7}$cm/s \cite{Kortus}. Hence, the upper limit for elastic mean
free path $l=\rho l/\rho_0$ for our film is about 3\,nm. In this
case, according to Eq.\,(1), the condition $d<l$ is fulfilled for
PC with $R>40\,\Omega $ or for lower resistance supposing multiple
contacts in parallel.

We have utilized generally used Blonder-Tinkham-Klapwijk equations
\cite{BTK82} describing $I-V$ characteristic of ballistic N-c-S
metallic junctions (here N is normal metal, c is constriction and
S is superconductor) by accounting for the processes of Andreev
reflection. At finite barrier strength at the N-S interface
characterized by parameter $Z\neq0$ and $T\ll T_c$, the theory
gives the d$V/$d$I$ curves with minima at $V\simeq \pm\Delta /$e.
To get the correct $\Delta $, the fit of the measured curves to
the theory should be done. The additional smearing of d$V/$d$I$
curves due to, e. g., broadening of the quasiparticle DOS in the
superconductor can be taken into account by including  parameter
$\Gamma$ \cite{Dynes}.

In the case of curves with double gap structure we calculated,
according to the theory \cite{BTK82}, the sum of two differential
conductances d$I/$d$V$ with the weight $w$ for the larger gap and,
correspondingly, with $(1-w)$ for the smaller one. After this, we
have transformed d$I/$d$V$ into d$V$/d$I$ to compare with the
measured dependences. The best fit was achieved, as a rule, by
using its own values of $Z$ and $\Gamma$ for large and small gap.
It is acceptable if we suppose that we have a number of
microconstrictions with various $Z$ in the region of mechanical
touch. It is worthy to note that, with increasing of weight
factor, the difference between $Z$ and $\Gamma$ values for large
and small gap becomes smaller or even vanishes for some PCs.

\textit{Results and discussion}. Approximately one half (44 of
total 91) of analyzed raw d$V$/d$I$ vs. $V$ curves show visible
two-gap structure, although, in most cases with shallow features
corresponding to a larger gap. The samples of some d$V/$d$I$
curves taken at 4.2\,K $\ll T_c$ with double-gap structure, along
with calculated curves, are shown in Fig.\,1. In spite of a number
of fitting parameters ($\Delta$, $\Gamma$, $Z$, $w$ ) for curves
with pronounced (or at least visible as shown in Fig.\,1) double
gap features determined $\Delta_L$ and $\Delta_S$ are robust as to
fitting procedure.

It turns out that histogram of gaps distribution built on the
basis of fitting of 44 spectra (see Fig.\,2a) has two
well-separated and quite narrow (especially for the small gap)
maxima.

The double-gap distribution is in qualitative agreement with the
distribution of gap values over the Fermi surface recently
calculated in \cite{Choi} (see Fig.\,2c). The main difference is
that theoretical distribution for lower gap is wider and has a
dominant maxima around 1.6 meV. This discrepancy can be resolved
when we draw attention that we have measured curves with double
gap structure for contacts that is predominantly along the c-axis.
In this case, according to \cite{Choi}, gap values along c-axis
spreads between 2 and 3 meV. The c-axis directionality of our
measurements is, apparently, the main reason of a shallow large
gap structure in d$V$/d$I$, because large gap dominates in the
"a-b" plane \cite{Choi}.

It should be mentioned that two very different order parameters
exist only in the clean limit $l\gg 2\pi\xi$. Since in our case
$l$ has upper limit in 3 nm and the coherence length $\xi\sim$5 nm
\cite{Finnemore}, the observation of two gaps is in line with our
supposition that in the PC area there are small grains with a much
larger mean free path. Indeed, SEM image of MgB$_2$ films
\cite{Bugos} shows that the film is granular with 100-500 nm large
grains. Therefore, in the area of mechanical contact there are
some amount of small metallic bridges, perhaps, with slightly
different crystallographic orientation being in parallel.

The single gap $\Delta$ is seen for the dirty limit \footnote{On
d$V/$d$I$ of "edge" contacts (a total of 11 curves) only single
gap structure was observed, probably due to the deterioration of
the film edge after breaking.} and is average of small and large
gaps with some weights. If we assume that this weight has some
relation to the weight $w$ used in the fitting procedure, then,
admittedly, $\Delta \simeq w\,\Delta_L +(1-w)\Delta_S$=3.4\,meV by
using upper limit $w\simeq$0.2 (see Fig.\,3). This agrees with the
position of the maximum of single-gap distribution at 3.5\,meV
(see. Fig.\,2b). By the way, according to the calculation in
\cite{Brink}, a large amount of impurity scattering will cause the
gaps to converge to $\Delta \simeq$4.1 meV.

Therefore the superconducting properties of this compound can be
strongly influenced by nonmagnetic defects and impurities, which
seem to have a great impact also on the scattering of gap value(s)
given by different authors.

As to $w$ factor it is hardly to see in Fig.\,3  its dependence on
$R_N$ or PC size, which one would expect if small gap reflects a
degraded surface or large gap is a result of surface states
\cite{Servedio}.

Table II shows double gap values given by different authors. A
quite good correspondence  between our results and data of other
authors carried out on different types of MgB$_2$ samples is
evident. In our case averaged over 44 curves, the ratio of the
larger gap to the lower one 2.85$\pm$0.15 is close to the
theoretical value 3:1 \cite{Liu}.

%\section{Conclusion}
\textit{In conclusion}. We have analyzed d$V$/d$I$ point-contact
spectra of MgB$_{2}$ with clear single- and double-gap structure.
The observed distinct maxima in the double gap distribution which
is consistent with theoretical calculations \cite{Choi} ruled out
surface or multiphase origin of gap structure and testify about
intrinsic superconducting double-gap state in MgB$_2$. The
averaged gap value ratio turned out to be in accordance with the
theoretically predicted ratio 1:3 \cite{Liu}.

%\section*{Acknowledgements}
\textit{Acknowledgments} The work in Ukraine was supported by the
State Foundation of Fundamental Research, Grant $\Phi$7/528-2001.
The work at Postech was supported by the Ministry of Science and
Technology of Korea through the Creative Research Initiative
Program. IKY is grateful to Forschungszentrum Karlsruhe for
hospitality.

\newpage

%FIGURE CAPTIONS

\begin{figure}
\includegraphics[width=7.5cm,angle=0]{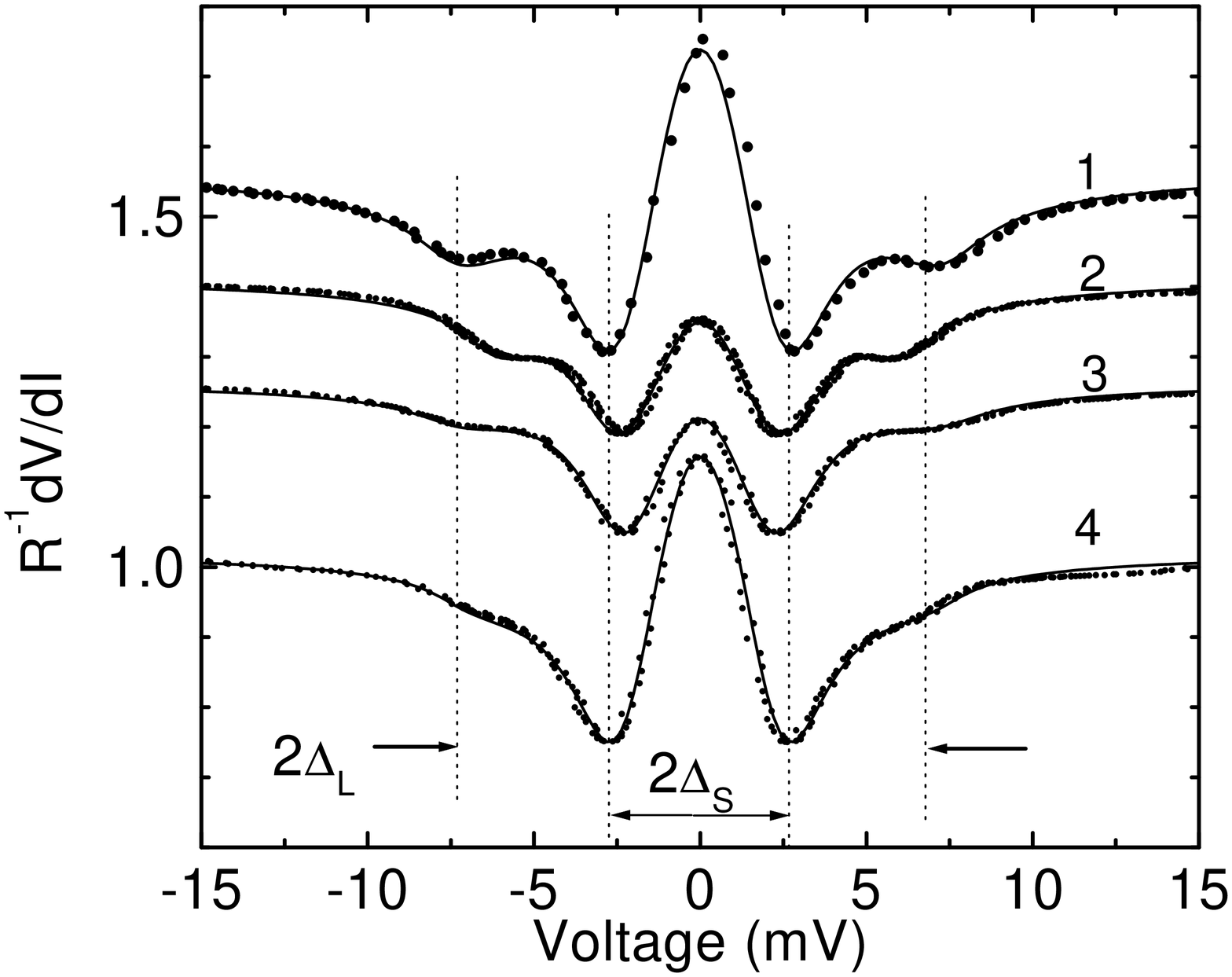}
\caption[] {Figure 1.\\ Reduced differential resistance
$R_N^{-1}$d$V$/d$I$ vs. $V$ measured at T = 4.2 K for four
MgB$_2$-Ag contacts with double gap structure (symbols). Thin
lines are theoretical dependences calculated with  parameters
given in the Table I. The curves (1-3) are vertically offset for
clarity. Vertical dashed lines show approximately position of
large $\Delta_L$ and small $\Delta_S$ gaps. Experimental curves
are taken nominally in c-directions} \label{fig1}
\end{figure}

\begin{figure}
\includegraphics[width=7cm,angle=0]{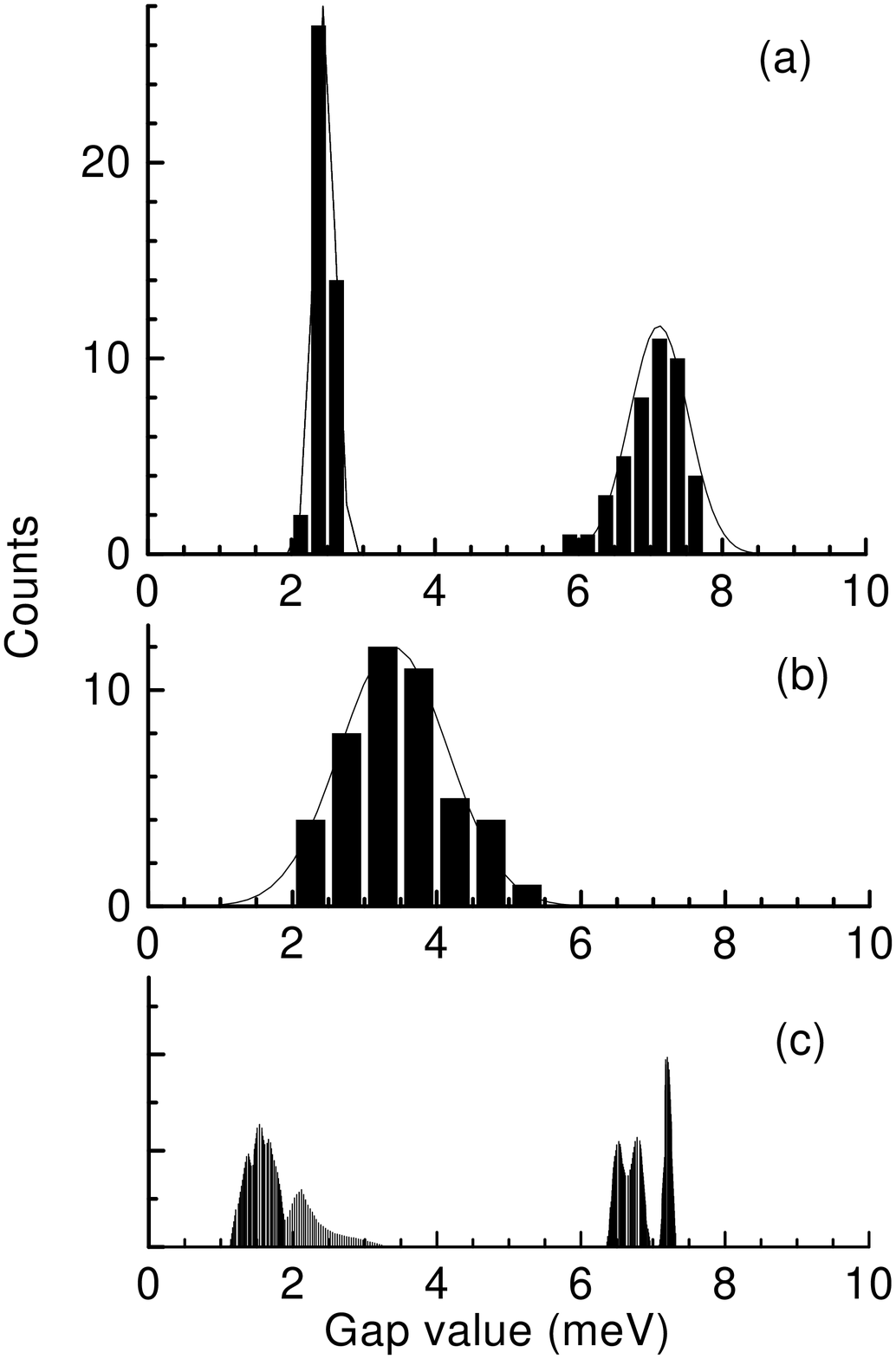}
\caption[] {Figure 2.\\ The superconducting energy-gap
distribution in c-axis-oriented  MgB$_{2}$ thin film in the case:
(a) double gap and (b) single gap. Thin lines show Gaussian fit
with maxima at (a) 2.45 and 7 meV and (b) 3.5 meV. The histogram
window of 0.25 meV for (a) and 0.5 meV for (b) is chosen to get
the most close to normal (Gaussian) distribution. (c) Distribution
of gap values over the Fermi surface calculated in \cite{Choi}}
\label{fig2}
\end{figure}

\begin{figure}
\includegraphics[width=7cm,angle=0]{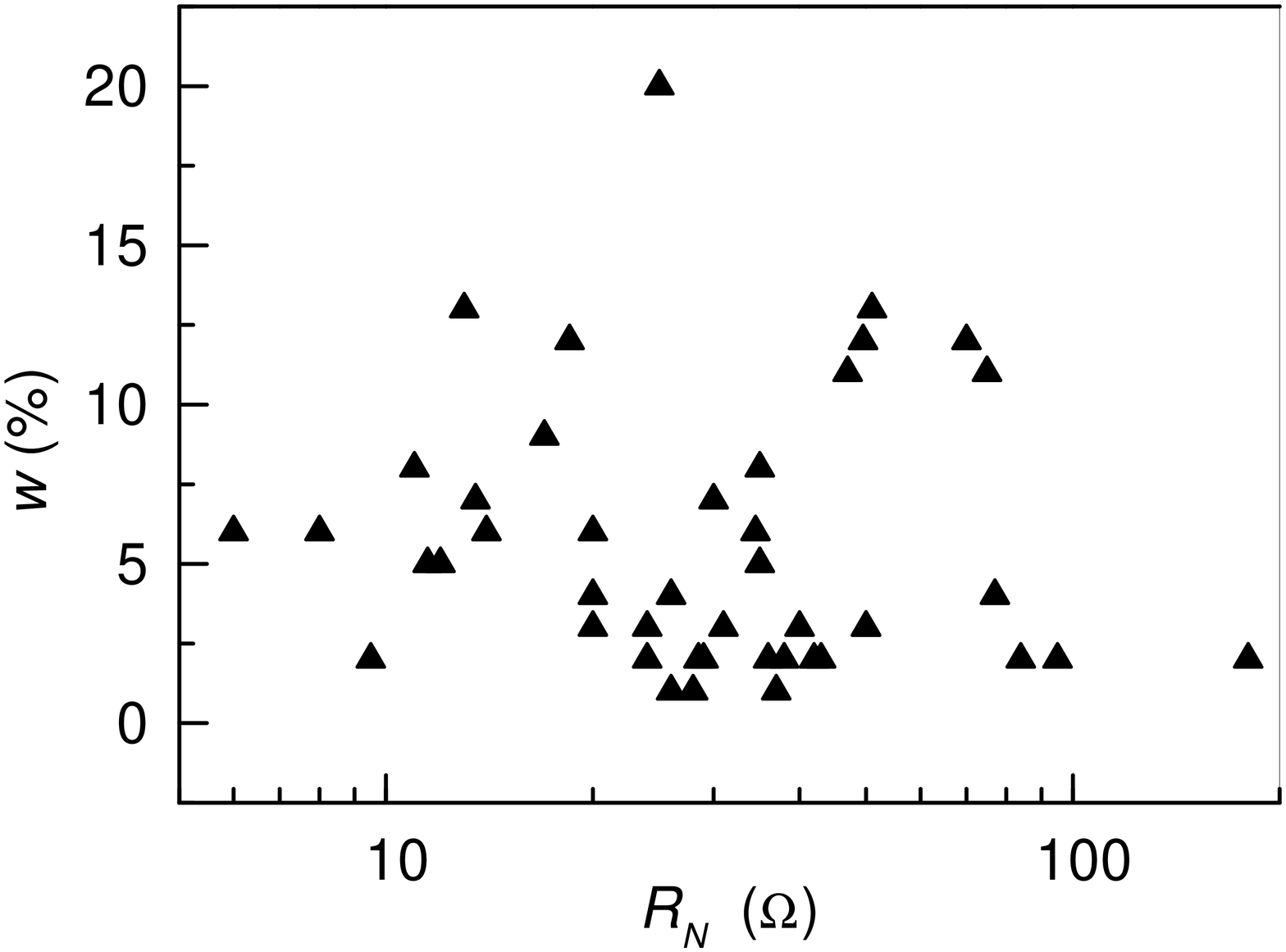}
\caption[] {Figure 3.\\  Dependence of the weight factor $w$ on
the point-contact resistance} \label{fig3}
\end{figure}

\newpage

%TABLES

\begin{table}
\caption{Table 1. Fitting parameters for curves presented in
Fig.\,1.}
%\begin{ruledtabular}
\begin{tabular}{ccccc}
\hline
 Parameters& Curve 1 & Curve 2 & Curve3 & Curve 4\\
\hline
R$_N$, $\Omega$ & 47 &  35 &  20 & 34\\
$\Delta_L$, meV & 7.4 & 6.25 & 7.35& 7.3\\
$\Delta_S$, meV & 2.6 & 2.54 & 2.4& 2.6\\
$w$-factor& 0.11 & 0.08 & 0.07& 0.06\\
Z$_L$ & 0.7 & 0.71 & 0.63& 0.21\\
Z$_S$ & 0.75 & 0.55 & 0.56& 0.76\\
$\Gamma_L$, meV & 0.4 & 0.1 & 0.55& 0\\
$\Gamma_S$, meV & 0.5 & 0.54 & 0.38& 0.3\\
\hline
\end{tabular}\label{tab1}
%\end{ruledtabular}
\end{table}

\begin{table}
\caption{Table 2. Gap values in MgB$_2$ measured by point-contact
(PCS) or tunneling spectroscopy (TS). }
%\begin{ruledtabular}
\begin{tabular}{ccccc}
\hline
 Method&Sample& $\Delta_S$, meV & $\Delta_L$, meV & Refs.\\
\hline PCS & Film &  $2.45\pm 0.15$ &  $7.0\pm 0.4$ & This pap.\\
PCS & Film & $2.3\pm 0.3$ & $6.2\pm 0.7$& \cite{Bugos}\\ PCS &
Grain & 2.8 & 7& \cite{Szabo}\\ TS & Granular & 3.9 & 7.5&\cite{Giubileo1}\\
TS & 50$\mu$ crys. & 3.8 & 7.8&\cite{Giubileo2}\\
TS & polycrys. & 1.75 & 8.2&\cite{Badr}\\
\hline
\end{tabular}\label{tab2}
%\end{ruledtabular}
\end{table}

\end{document}